\documentclass[dvips,twoside,fleqn]{article}
\usepackage{epsfig,espcrc2}

\newcommand{\AmS}{{\protect\the\textfont2
  A\kern-.1667em\lower.5ex\hbox{M}\kern-.125emS}}

\newcommand{\GS}{\gamma_{\rm str}}
\newcommand{\BE}{\begin{equation}}
\newcommand{\EE}{\end{equation}}

\newcommand{\la}{\lambda}

\title{
\vspace{-5.0cm}
\begin{flushright}
{\normalsize FUB-HEP 96-10}\\
{\normalsize \today}\\
\end{flushright}
\vspace*{3.2cm}
Fixed versus random triangulations in 2D simplicial Regge calculus 
\thanks{Work supported in part by the EEC under contract No. ERBCHRXCT930343
and by the NVV grant bvpf01.}}

\author{Christian Holm\address{Institut f\"ur Theoretische Physik,
        Freie Universit\"at Berlin, 14195 Berlin, Germany} and 
        Wolfhard Janke$^{\rm a\,\!}$\address{Institut f\"ur Physik,
Johannes Gutenberg-Universit\"at Mainz, 
55099 Mainz, Germany}
\thanks{WJ thanks the DFG for a Heisenberg fellowship.}} 

\begin{document}
\begin{abstract}

We study 2D quantum gravity on spherical 
topologies using the Regge calculus approach with the $dl/l$ measure. 
Instead of a fixed non-regular triangulation which has been used before, we
study for each system size four different random triangulations, 
which are obtained according to the standard Voronoi-Delaunay procedure. 
We compare both approaches quantitatively and show
that the difference in the expectation value of $R^2$
between the fixed and the random
triangulation depends on the lattice size and the surface area 
$A$. 
We also try again to measure the string susceptibility exponents 
through a finite-size scaling
Ansatz in the expectation value of an added $R^2$ interaction term in an
approach where $A$ is held fixed. 
The string susceptibility exponent $\gamma_{\rm str}'$ is shown to agree with
theoretical predictions
for the sphere, whereas the estimate for $\gamma_{\rm str}$ appears to be
too negative.

\end{abstract}

\maketitle

                  \section{INTRODUCTION}
In the past few years Regge calculus has been extensively used in the study of 
quantum gravity \cite{q-regge}. In its usual form one
studies regular simplicial triangulations of manifolds of a given topology, 
which are
mostly hypertori. In this work we will investigate 
two-dimensional quantum gravity, 
where analytic calculations
have shown that the internal fractal structure of the manifold depends very
sensitively on the global topology. One universal quantity is
the string susceptibility exponent
$\gamma _{\rm str}$, which is the sub-dominant correction to the large area
behavior
of the partition function 
$Z(A) \propto A^{\gamma_{\rm str} -3} e^{- \lambda _R A}$,
where $\lambda _R $ denotes the renormalized cosmological constant.
The exponent $\gamma_{\rm str}$ depends on the genus $g$ of the surface
through the KPZ formula 
$\gamma_{\rm str} = 2 - \frac{5}{2} (1-g)$ \cite{gamma_string}.

For the torus ($g = 1$) the Regge approach with the $dl/l$ measure gives 
compatible results. 
For the sphere ($g = 0$) and topologies of higher gender, however, 
the situation is still unclear \cite{hamber91,bock,bock_proc,hj95a,hj95b}.
One potential problem is 
that for the sphere only very small regular triangulations exist, such
as
the tetrahedron, the octahedron, and the icosahedron. 
In order to obtain 
triangulations on larger lattices,
one either has to use non-regular triangulations, with a few special vertices
which might spoil the finite-size scaling (FSS) behavior, 
or to resort to random
triangulations. In this work we use Monte Carlo (MC) simulations to 
study in detail random triangulations of a
sphere and
compare the results with our earlier results 
obtained by using the triangulated surface of a cube as spherical 
lattice \cite{hj95b}.
In addition we present estimates of $\GS$ and the related 
exponent $\GS'$ \cite{kawai_r2} on the random lattices, 
employing the novel FSS method of Ref.~\cite{hj95a}.
            \section{MODEL}
For a measurement of the string
susceptibilities in Regge calculus one needs to introduce a curvature
square term in the action, 
and then deduces from its expectation value an estimate of $\GS$ and $\GS'$ 
through FSS analyses. We therefore considered
the partition function
\begin{equation}
Z(A)=\!\!\int{\!\!{\cal D} \mu (q) e^{-\!\sum _i
{(\lambda A_i + a R_i^2)}} \delta (\sum _i A_i-A)},
\label{eq:1}
\end{equation}
where $R _i ^2 = \delta _i ^2/A_i$ denotes the local squared curvatures.
The $A_i$ are barycentric areas and
$\delta _i = 2 \pi - \sum_{t \supset i} {\theta _i (t)}$ are the
deficit angles, with $\theta_i (t)$ being the dihedral angle at vertex
$i$. The dynamical degrees of freedom are the squared link lengths, 
$q=l^2$, which stand in a linear relation to the 
components of the metric tensor $g$. We used the simple scale invariant 
``computer measure''
${\cal D}\mu (q) = \left[ \prod _{\langle ij \rangle} \frac{dq_{ij}}{q_{ij}} 
\right] F_\epsilon(\{q_{ij}\})$.
The notation is identical to that used in Ref.~\cite{hj94a}.

The only dynamical
term is the $R^2$-interaction, because we held $A$ fixed during the update. 
The $R^2$ coupling constant $a$   
sets a length scale of $\sqrt{a}$, and $\hat A := A/a$
can be used to distinguish between the cases of 
weak $R^2$-gravity ($\hat A \gg 1$), where
the KPZ scaling
is recovered, and strong $R^2$-gravity 
($\hat A \ll 1$)
where it was found \cite{kawai_r2}, that
$Z(A) \propto A^{\gamma_{\rm str}' -3} e^{-S_c/\hat A} e^{- \lambda _R A -
b \hat A }$,
with the classical action
$S_c = 16 \pi ^2(1-g)^2$, some constant $b$, and 
$\gamma_{\rm str}' = 2 - 2(1-g)$.

As global
lattice topology we used  
a randomly triangulated sphere constructed according to the
Voronoi-Delaunay procedure, see Fig.~\ref{fig1} for a sample lattice. 
In this way we can control the influence of
non-regular triangulations.
For spherical topologies we have the relations
$N_0 - 2 =  N_2/2$, $N_0 - 2 = N_1/3$, and 
$2N_1=3N_2$, where $N_0, N_1$, and $N_2$ 
denote the number of sites, links and triangles, respectively.
%
                   \section{FINITE-SIZE SCALING}
%
All previously used methods to extract $\GS$ \cite{hamber91,bock,bock_proc} 
are plagued by inconsistencies, as has been discussed in our earlier 
work \cite{hj95b}.
We suggested a new approach which is in spirit much closer to the continuum
analysis of Ref.~\cite{kawai_r2} as it uses $\hat A$ as the 
distinguishing parameter
between weak and strong $R^2$ gravity.
The dimensionless expectation value 
$\hat R^2 := a\langle \sum _i R^2 _i \rangle$ 
can be shown to depend only on $N_2$ and the dimensionless parameter $\hat A$.
Sending $N_2 \rightarrow \infty$ one expects
$\hat R^2(\hat A,\infty)$
to be expandable
in a power series, whose first three terms read as
\BE
\hat R^2 (\hat A) = \dots + b_0 \hat A + b_1 + b_2 / \hat A + \dots,
\label{eqn:R-infty}
\EE
where $b_0 = -(\la _R - \la )a, b_1 = \GS -
2$ for
$\hat A \gg 1$, and
$b_0 = -b - (\la _R - \la )a, b_1 = \GS ' - 2, b_2 = S_c$ 
for $\hat A \ll 1$.

Expanding $\hat R^2(\hat A,N_2)$ at constant $\hat A$ we obtain
\BE
\hat R^2(\hat A, N_2) = N_2d_0(\hat A) +
d_1(\hat A) + d_2(\hat A)/N_2 + \dots
\label{eq:our_N_sc}
\EE 
The next step is to expand the coefficients $d_i$ as a power series
in $\hat A$.
The coefficient $d_1$ carries all the necessary information
to extract the string susceptibilities.
A comparison with (\ref{eqn:R-infty}) yields
\begin{equation}
d_1(\hat A) =  b_0 \hat A  + \GS - 2 + {\cal O}(1/\hat A)
\label{eq:gg1}
\end{equation}
for $\hat A \gg 1$ and 
\begin{equation}
d_1(\hat A) = S_c/\hat A  + \GS' - 2 +  b_0 \hat A  + {\cal O}(\hat A ^2)
\label{eq:ll1}
\EE
for $\hat A \ll 1$.
If we plot $d_1$ versus $\hat A$ we expect to see a linear behavior for very
large $\hat A$, and a divergent behavior for small $\hat A$, from which we
can extract $\GS$ as well as $\GS'$.
Because $\hat R^2$ in (\ref{eq:our_N_sc}) 
becomes infinite in the continuum limit $N_2 \longrightarrow \infty$, 
it was suggested
to add a non-scale invariant part $q_{ij}^\alpha$ to the measure and fine-tune
$\alpha$ such that it cancels the divergent term $d_0$. However, a trial
simulation showed within error bars no change in the relevant coefficient
$d_1(\hat A)$ \cite{hj95b}.
\begin{figure}[th]
\begin{center}
 \epsfig{file=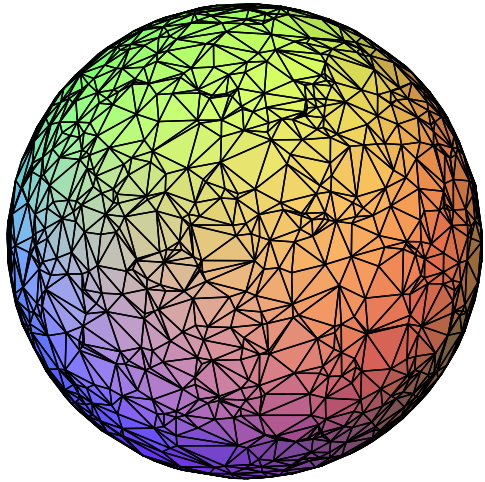,width=55mm}
\end{center}
\vspace{-0.7cm} \caption{Randomly triangulated sphere with $N_0=1500$.}
\label{fig1}
\end{figure}
%
%
                     \section{SIMULATION}
%
For each lattice size we generated four different randomly
triangulated
spheres constructed according to
the Voronoi-Delaunay procedure.
Usually, the size of the lattices varied from 218 up to 17\,498 lattice sites,
corresponding to 648 -- 52\,488 link degrees of freedom, 
or 432 -- 34\,992 triangles.
To update the links we used a standard multi-hit
Metropolis algorithm with a hit rate ranging from 1,\dots ,3.
The area was kept fixed with a value of $\hat A$ in the range of 
7 -- 1\,800.
For each run on the four copies we recorded about 
20\,000 -- 50\,000 measurements 
of the curvature square 
$R^2 := \sum _i R^2_i$ on every second to fourth MC sweep. 
The statistical errors for each copy were computed using standard jack-knife
errors on the basis of 20 blocks. The integrated autocorrelation time
$\tau _{R^2}$ of $R^2$ was usually in the range of 5 -- 10. As final error in
the average of the four copies we used the standard root mean square deviation.
                     \section{RESULTS}
We first begin with a comparison of the raw data for 
$\hat R^2 /N_0$ obtained on the randomly triangulated sphere and our earlier
data produced on the surface of a cube, see Fig.~\ref{fig2}. The difference in
$\hat R^2$ depends, as could be expected, on the lattice size, such that the
difference between the two triangulations decreases as $N_0$ increases. This is
also true for the difference in $\hat R^2$ between different copies of the
random triangulations. Noteworthy is, that for the larger system sizes the
copies assume almost the same value within their statistical error.
\begin{figure}[h]
{\hspace{-0.7cm} \epsfig{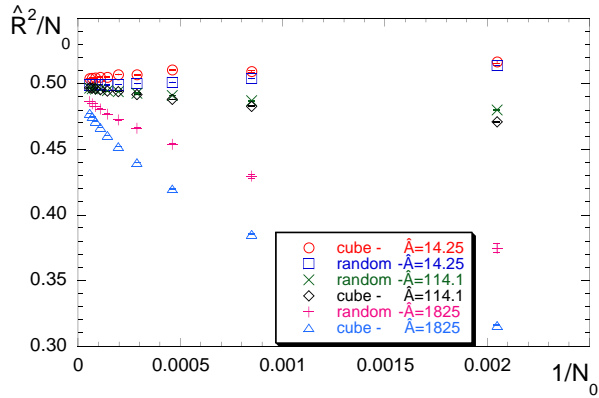}}
\vspace{-1.5cm} \caption{Comparison of raw data for $\hat R^2 /N_0$ obtained
  for various 
  lattice sizes on the non-regular cube and the randomly triangulated sphere.}
\label{fig2}
\end{figure}

However, the difference in $\hat R^2$
also depends on $\hat A$ such that for small $\hat A$ the value of $\hat
R^2$ is larger on the cube, whereas for large $\hat A$ the value of $\hat R^2$
is smaller on the cube than on the randomly triangulated sphere. This effect is
not visible among the four copies of the random triangulations.
Strictly
speaking this shows that $Z$ of eq.~(\ref{eq:1}) and $\hat R^2$ of
eq.~(\ref{eq:our_N_sc}) depend also on the way the manifold is triangulated,
i.e. the incidence matrix. Especially for large $\hat A$, i.e. the region
which determines $\GS$, the values for $d_1$
depend on the triangulation, and one can expect similar values only for very
large lattices. 

We used linear two-parameter fits in eq.~(\ref{eq:our_N_sc}) 
to extract from our raw data for $\hat R^2(\hat A, N_2)$ the
values $d_1(\hat A)$. A number of data points on the smaller lattices had to be
discarded until the fit reaches a sufficiently high quality. 
All data points for $d_1$ obtained in this way are shown 
in Fig.~\ref{fig3}.
\begin{figure}[h]
{\hspace{-0.7cm} \epsfig{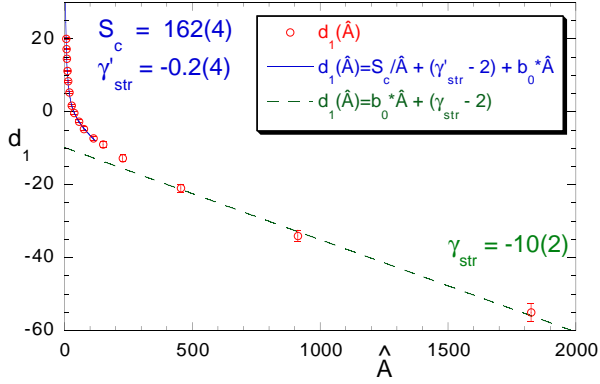}}
\vspace{-1.2cm} 
\caption{Results for $d_1(\hat A)$ vs.\ $\hat A$ 
together with fits in the regimes 
$\hat A < 120$ and $\hat  A > 400$ yielding estimates for 
$\GS'$ and $\GS$.}
\label{fig3}
\end{figure}

In the next step we fitted $d_1(\hat A)$
for small values of $\hat A < 120$ according to the Ansatz
(\ref{eq:ll1}), which yields $S_c = 162(4)$ and $\GS' = -0.2(4)$ with a total 
$\chi ^2
= 2.4$, see Fig.~\ref{fig4}.
This result is perfectly compatible with the theoretical prediction 
$S_c = 16\pi ^2 \approx  158$ and $\GS' = 0$.
\begin{figure}[h]
{\hspace{-0.7cm} \epsfig{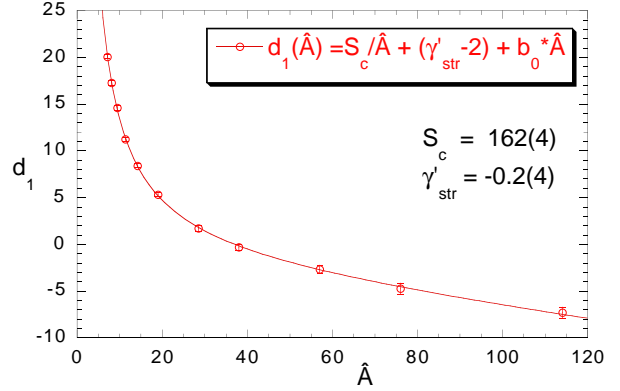}}
\vspace{-1.2cm} \caption{Blow-up of Fig.~3 for $\hat A < 120$ yielding an
  estimate for 
$\GS'$.}
\label{fig4}
\end{figure}

For large $\hat A$ we can employ Ansatz (\ref{eq:gg1}). However, there are 
only three data points with a sufficiently large $\hat A$ available. The
linear fit yields $\GS = -10(2)$ with $\chi ^2 = 0.8$.
It is not clear, however,
if we are already in the asymptotic regime, which might set in at much larger
$\hat A$. A further difficulty is that one is not interested in the slope, but
in the intersection of the fit with the $y-$axis, which is far from the
location of the points used in the fit. 
The systematic uncertainty on our estimate of
$\GS$ is therefore 
hard to estimate, but it is interesting to note that our value is too
negative, which is just opposite to what has been claimed in Ref. \cite{bock}
by using a different FSS method.
                   \section{CONCLUSIONS}
%
Random triangulations appear to be a good alternative for topologies where no
large regular triangulations exist. 
They show good scaling behavior, and the differences between different copies
of the same area
decrease as the system size increases. In this way they can
provide a ``typical'' lattice for the evaluation of expectation values with 
the partition function of eq.~(\ref{eq:1}). 

The quantitative difference in $\hat R^2$ between the non-regular triangulation
and the random triangulation of the sphere depends on both, $\hat A$ and
$N_0$. In this way one will obtain on the usually used system sizes different
values of $d_1(\hat A)$. The difference seems to be negligible for small
values of $\hat A$, so that $\GS'$ can be consistently obtained on both, the
cube and the randomly triangulated sphere. However,
the difference becomes important for large values
of $\hat A$, and is thus a potential problem for the determination of $\GS$.

Our FSS method of fitting at constant values of $\hat A$ 
gives results for $\GS'$ which are compatible with the
theoretical prediction. In contrast to Ref. \cite{bock_proc}
we employ a consistent FSS scheme and also much larger lattices. It
would be interesting to test if contrary to \cite{bock_proc} also for
topologies of higher gender the theoretical expectations 
for $\GS'$ can be confirmed.

Due to the fewer data points, only a crude estimate for $\GS$ could be
obtained which, however, appeared to be too negative compared to the KPZ
theory. 
This is exactly opposite to what has been found in \cite{bock} with a
different FSS Ansatz. We
attribute this discrepancy to their method, which in our opinion \cite{hj95b},
bears conceptual problems  for large values of $\hat A$. 
It is yet unclear, if our system sizes are already in the
asymptotic scaling regime, so that the potential danger of systematic errors 
is still very large.
\vspace{-0.2cm}
            
\end{document}